\documentclass{vgtc}                          

\ifpdf
  \pdfoutput=1\relax                   
  \usepackage{graphicx}                
  \DeclareGraphicsExtensions{.pdf,.png,.jpg,.jpeg} 
\else
  \usepackage{graphicx}                
  \DeclareGraphicsExtensions{.eps}     
\fi%

\graphicspath{{figures/}{pictures/}{images/}{./}} 

\usepackage{microtype}                 
\PassOptionsToPackage{warn}{textcomp}  
\usepackage{textcomp}                  
\usepackage{mathptmx}                  
\usepackage{times}                     
\usepackage{cite}                      
\usepackage{tabu}                      
\usepackage{array, booktabs, tabularx}

\onlineid{0}

\vgtccategory{Research}

\vgtcinsertpkg

\title{Road Accidents in the UK (Analysis and Visualization)}

\author{Anjul K. Tyagi, Ayush Kumar, Anshul Gandhi, Klaus Mueller\thanks{e-mail: \{aktyagi, aykumar, anshul, mueller\}@cs.stonybrook.edu}}
\affiliation{\scriptsize Department of Computer Science, Stony Brook University, New York}

\teaser{
 \includegraphics[width=1\textwidth]{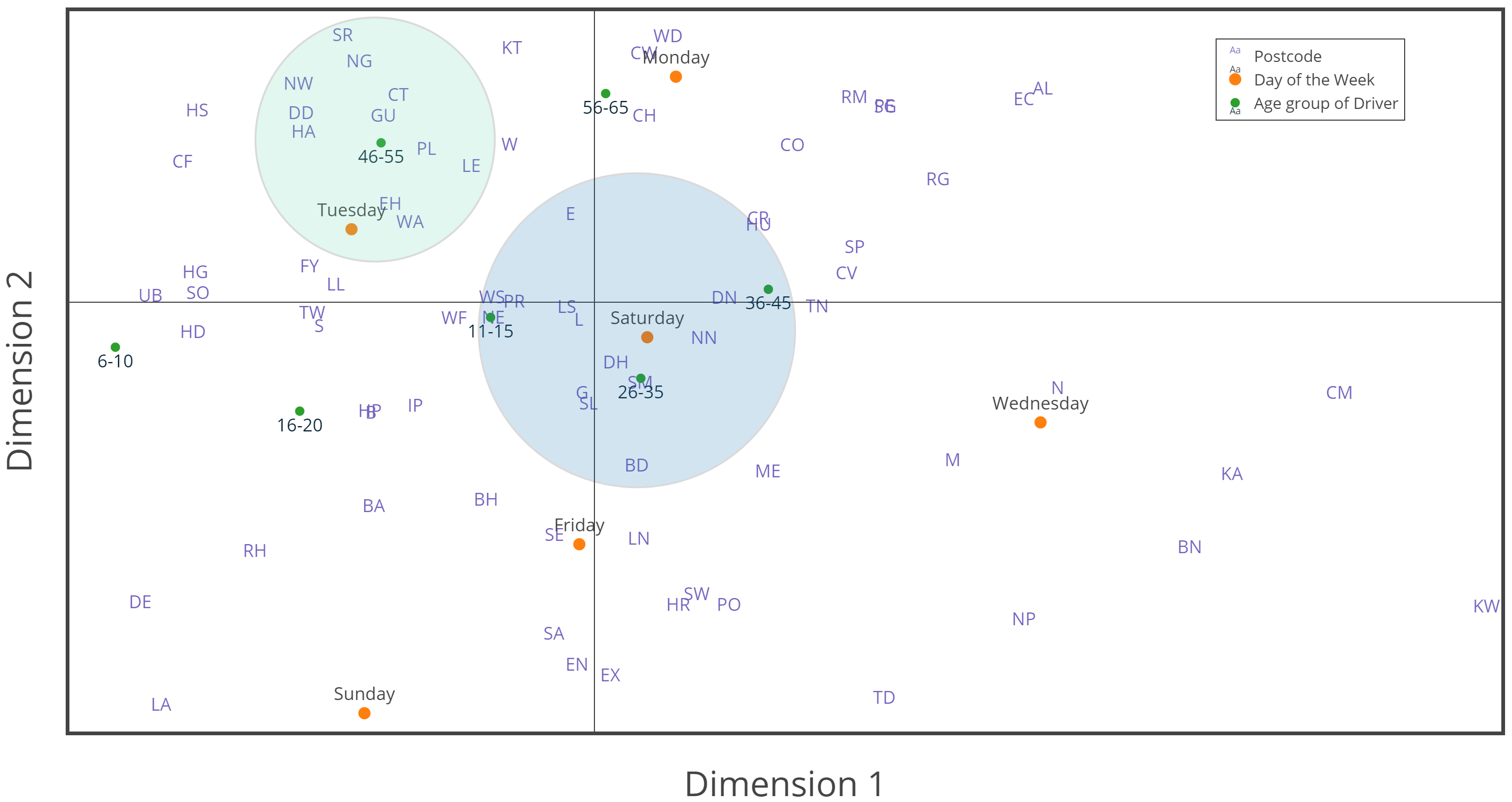}
 \caption{Multiple Correspondence Analysis of road accidents based on \textbf{Postcode, Day of the week and Age of Drivers}. Liverpool (Postcode: L) has more accidents on Saturdays with majority drivers lying in the age group of 26 to 35 years.}
 \label{mca}
}

\abstract{Analysis of road accidents is crucial to understand the factors involved and their impact. Accidents usually involve multiple variables like time, weather conditions, age of driver etc. and hence it is challenging to analyze the data. To solve this problem, we use Multiple Correspondence Analysis (MCA) to first, filter out the most number of variables which can be visualized effectively in two dimensions and then study the correlations among these variables in a two dimensional scatter plot. Other variables, for which MCA cannot capture ample variance in the projected dimensions, we use hypothesis testing and time series analysis for the study.
} 

\CCScatlist{Multiple Correspondence Analysis---Dimension Reduction---Time Series Analysis---Hypothesis Testing}

\begin{document}



\maketitle

\section{Introduction}
Analysis of road accidents data can reveal various hidden facts. Accident datasets are high dimensional in nature and techniques like MDS and PCA can be used to project the data in lower dimensions for visualization. However, these techniques don't preserve the correlation among variables. Instead, Multiple Correspondence Analysis\cite{abdi2007multiple} (MCA) can be used to visualize and correlate between variables from the high dimensional data in two dimensions. It also gives the discrimination measure of how correctly each variable from the dataset is represented in lower dimensions. We use this measure to effectively visualize some variables from the dataset. For other variables which can't be correctly visualized by MCA, we use hypothesis testing and time series analysis to get some further insights.    

\section{Dataset}
The dataset is taken from Kaggle\cite{kaggle} and it contains the details of every recorded accident in the UK from 2005 till 2015. The full dataset is divided into three major categories i.e. accident information, casualty information, and vehicle information.

\section{Related Works}
Ljubic et al. \cite{ljubivc2002time} used time series analysis to study the accidents data in the UK. Sikdar et al. \cite{sikdar2017hypothesis} used hypothesis testing to study accidents data in India. However, our work uses data visualization to filter out a smaller set of features which can't be effectively visualized in lower dimensions.

\section{Approach}
\subsection{Discrimination measure over variables}
\begin{figure}[tb]
 \centering 
 \includegraphics[width=\columnwidth]{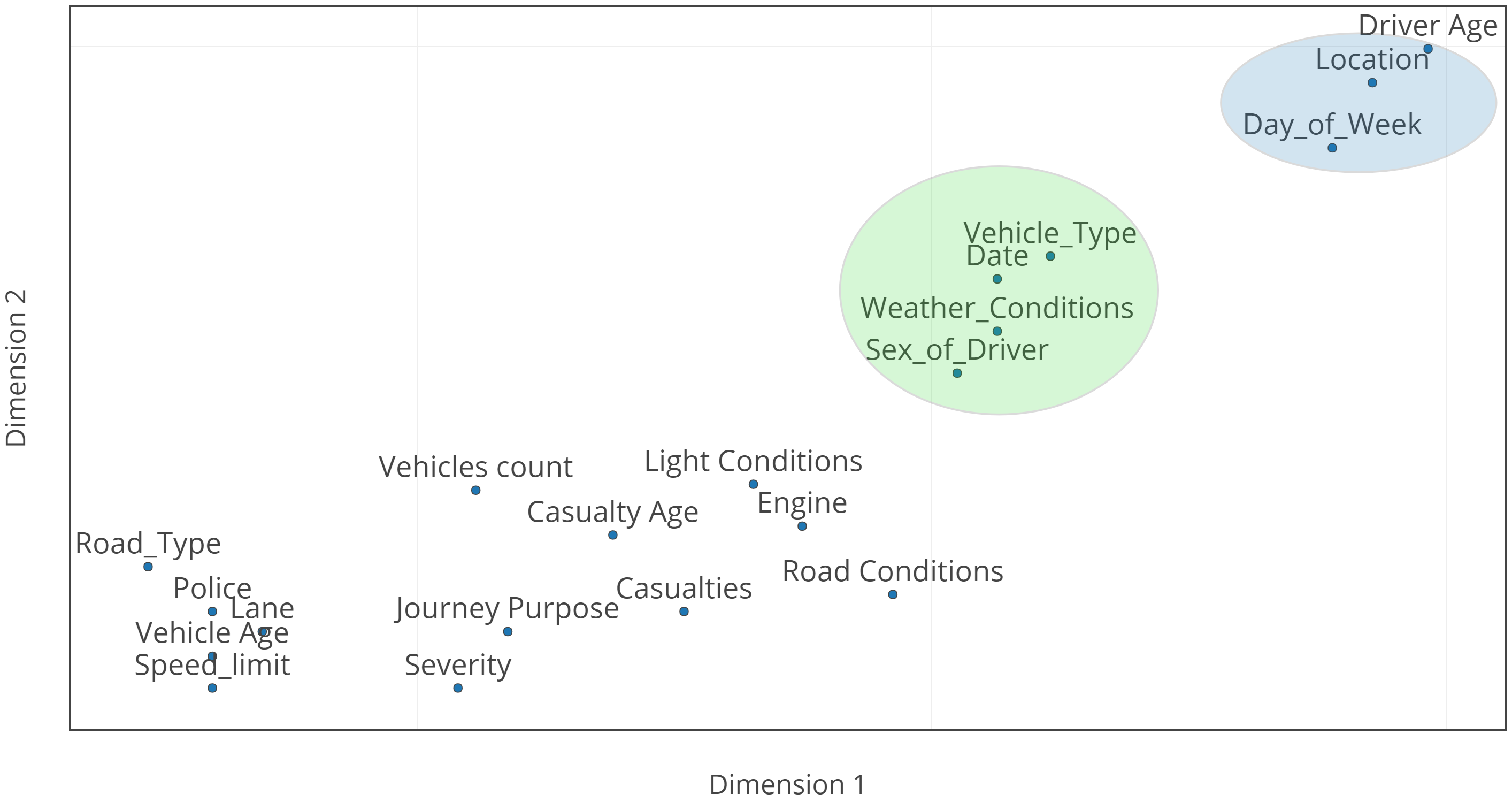}
 \caption{Discrimination Measure plot over variables in accidents data.}
 \label{fig:correlation}
\end{figure}

We use discrimination plot generated with MCA to see which variables can be represented accurately in two-dimensional visualizations of the dataset. As shown in Figure \ref{fig:correlation}, more the value of a variable along any dimension, easier it is to represent that variable along that dimension. As the circles in Figure  \ref{fig:correlation} show, the main variables which can be visualized using MCA are \textbf{Age of the driver, the location of an accident, day of the week}. Other variables like vehicle type, date, weather conditions and sex of the driver which cannot be accurately represented using MCA, we use hypothesis testing and time series analysis to analyze them.

\subsection{Multiple Correspondence Analysis variables plot}
We choose three variables which had high variance along Dimensions 1 and 2 from Discrimination Measure Analysis, namely \textbf{Location of the accident (Postcode), Day of the week and the Age group of the driver} and project them using MCA with all the features in a single plot to study the correlation. Figure\ref{mca} shows how each of these categories is related to others in the form of a scatter plot. Several insights obtained from this analysis are discussed in the results section.

\subsection{Analysis over other variables and important events in the UK}
Not all the variables can be efficiently represented in lower dimensions using MCA, hence further techniques to analyze data are required. Hypothesis testing can be used to further understand how the variables are related to each other. We used Welch's t-test statistic to study several hypothesis on this dataset. Furthermore, because this dataset is time-bound, we can make some predictions on the data using time series analysis. We applied autoregression on our dataset to analyze and predict the trend in accidents over the years. Results are discussed in the next section.

\begin{figure}[tb]
 \centering 
 \includegraphics[width=\columnwidth]{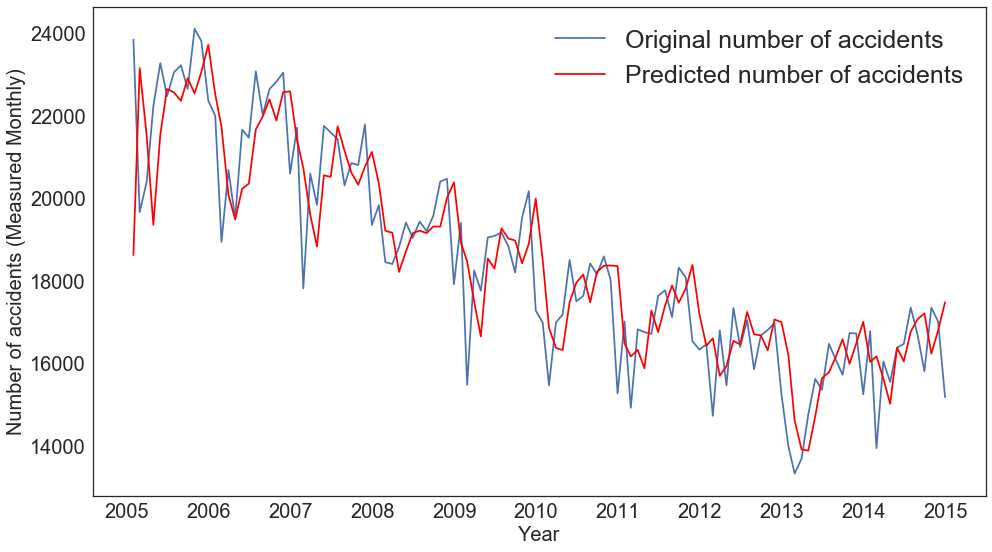}
 \caption{Prediction of monthly accidents using autoregression from 2005 till 2014.}
 \label{autoregression}
\end{figure}


\section{Results}
\subsection{MCA plot with postcode, day of the week and age group of driver (Figure \ref{mca})}

\begin{itemize}
\setlength\itemsep{0em}
\item The number of accidents on Sundays and Wednesdays is fewer than those on other days in any postcode.
\item Age groups 11-15 years, 26-35 years and 36-45 years have the similar number of accident records and the major day of accidents for these age groups is Saturday.
\item Warrington(WA) and Guildford(GU) have more accidents on Tuesdays and the most common age group of people causing accidents is 46 to 55 years.
\item Age group 6-10 years is responsible for a lesser number of accidents compared to other age groups.
\end{itemize}

\subsection{Hypothesis testing}
We found out that the number of accidents before, and during the London Summer Olympics remained same. Similarly, other interesting hypothesis were tested and are discussed in Table \ref{tab:hyp_results}.  

\subsection{Time Series Analysis}
Figure \ref{autoregression} shows the prediction of the number of monthly accidents over the years. We see that the number of accidents has decreased over the years. The prediction accuracy can be measured by the root mean square error value, which was 699.84.

\section{Conclusion}
In this paper, we combined visualization and data analysis techniques for the effective study of a dataset. We visualized the correlation between the location of the accident, day of the week and age of the drivers using MCA. Further, we studied other important features using hypothesis testing and predicted the trend in accidents using time series analysis. Future work will include more detailed analysis of the data using Machine Learning and other advanced visualization techniques.

\section{Acknowledgments}
This research was partially supported by NSF grant IIS 1527200 \& MSIT, Korea under the ICTCC Program (IITP-2017-R0346-16-1007).

\newcolumntype{C}{>{\arraybackslash}X}
\begin{table}[tb]
\caption{Results from hypothesis testing.}
\begin{tabularx}{\linewidth}[tb]{*{2}{C}}
  \toprule
   \textbf{Null Hypothesis} & \textbf{Result} \\
  \midrule
  Number of daily accidents in summer and winter are equal & 15 to 30 more daily accidents in summer\\ \hline 
  Number of daily accidents by young drivers (Age 18-25 years) and old drivers (Age 65-85 years) are equal & 85 to 89 more accidents by young people\\ \hline
  The number of daily accidents before and during the London Summer Olympics (2012) were same & Accept Null Hypothesis. P-value 0.197.\\ \hline
  Number of daily accidents in areas close to subway stations is same as other areas & 9 to 29 more accidents daily in areas close to subway stations.\\ \hline
  Males cause an equal number of daily accidents as females & 428 to 439 more accidents by males.\\
  \midrule
\end{tabularx}
 \label{tab:hyp_results}
\end{table}

\bibliographystyle{abbrv-doi}

\bibliography{template}
\end{document}